\documentclass{aa} 

\usepackage{natbib}
\bibpunct{(}{)}{;}{a}{}{,} 

\usepackage{graphicx}

\usepackage{amssymb,amsmath}
\usepackage{delarray}
\usepackage{mathrsfs}
\usepackage{xcolor}

\usepackage[varg]{txfonts}
\usepackage{float}

\definecolor{darkgreen}{rgb}{0.0, 0.60, 0.0}

\usepackage{soul} 

\begin{document} 

\title{The polarization angle in the wings of Ca~{\sc{i}} 4227: A new observable for diagnosing unresolved photospheric magnetic fields}

\author{Emilia Capozzi\inst{1,2}
\and
Ernest Alsina Ballester\inst{1,3,4}
\and
Luca Belluzzi\inst{1,5,6}
\and
Javier Trujillo Bueno\inst{3,4,7}
}

\institute{
Istituto Ricerche Solari (IRSOL), Universit\`a della Svizzera italiana, CH-6605 Locarno-Monti, Switzerland
\and 
Geneva Observatory, University of Geneva, CH-1290 Sauverny, Switzerland
\and
Instituto de Astrof\'isica de Canarias, E-38205 La Laguna, Tenerife, Spain
\and
Departamento de Astrof\'isica, Universidad de La Laguna, E-38206 La Laguna, Tenerife, Spain
\and 
Leibniz-Institut f\"ur Sonnenphysik (KIS), D-79104 Freiburg, Germany
\and
Euler Institute, Universit\`a della Svizzera italiana, CH-6900 Lugano, Switzerland
\and 
Consejo Superior de Investigaciones Cient\'ificas, Spain\\
\email{emilia.capozzi@irsol.usi.ch}
}

\titlerunning{New observable for unresolved magnetic field diagnostics}
\authorrunning{Capozzi et al.}

\abstract
{When observed in quiet regions close to the solar limb, many strong resonance lines show conspicuous linear polarization signals, produced by scattering processes (i.e., scattering polarization), with extended wing lobes.
Recent studies indicate that, contrary to what was previously 
believed, the wing lobes are sensitive to the presence of relatively weak longitudinal magnetic fields through magneto-optical (MO) effects.} 
{We theoretically investigate the sensitivity of the scattering polarization wings of the 
Ca~{\sc{i}} 4227\,{\AA} line to the MO effects, and we explore its diagnostic potential 
for inferring information on the longitudinal 
component of the photospheric magnetic field.} 
{We calculate the intensity and polarization profiles of the Ca~{\sc i} 4227\,{\AA} line by numerically 
solving the problem of the generation and transfer of polarized radiation under non-local thermodynamic 
equilibrium conditions in one-dimensional semi-empirical models of the solar atmosphere, taking into account 
the joint action of the Hanle, Zeeman, and MO effects.
We consider volume-filling magnetic fields as well as magnetic fields occupying a fraction of the  
resolution element.}
{In contrast to the circular polarization signals produced by the Zeeman effect,  
we find that the linear polarization angle in the scattering polarization wings of Ca~{\sc i} 4227\,
presents a clear sensitivity, through MO effects,
not only to the flux of the photospheric magnetic field, but also to the fraction of the resolution element that 
the magnetic field occupies.}
{We identify the linear polarization angle in the  
wings of strong resonance lines as a valuable observable 
for diagnosing unresolved magnetic fields.
Used in combination with observables that encode information on the magnetic flux and other properties 
of the observed atmospheric region (e.g., temperature and density), it can provide constraints on the filling factor of the magnetic field.}
{}


\maketitle

\section{\textbf{Introduction}}\label{intro}
The radiation escaping from the solar atmosphere encodes valuable information on its 
thermodynamic and magnetic properties. 
Strong resonance lines that show broad profiles in the solar spectrum are of particular 
diagnostic value; because their core and wing photons originate at significantly different heights, they encode information about extended regions of the solar atmosphere. 
When observed close to the solar limb, these lines generally present strong linear 
polarization signals 
arising from the scattering of anisotropic radiation 
\citep[i.e., scattering polarization;][]{LandiE}. 
A remarkable example is the Ca~{\sc i} 4227\,{\AA} line, which shows the 
largest scattering polarization signal in the visible part of the second solar 
spectrum \citep{Bruckner63,Stenflo80,Stenflo84,Gandorfer02}.
This signal presents a characteristic triplet-peak structure, consisting of a sharp peak 
in the core and broad lobes in the wings. 
It is well established that the scattering polarization 
wing lobes and line-core peak encode information on the thermodynamical structure of the photosphere and the lower chromosphere, respectively. 
Moreover, the latter is known to be sensitive to chromospheric magnetic fields
through the Hanle effect \citep{stenflo01}.
\begin{figure}[h!]
\centering
\includegraphics[width=.5\textwidth]{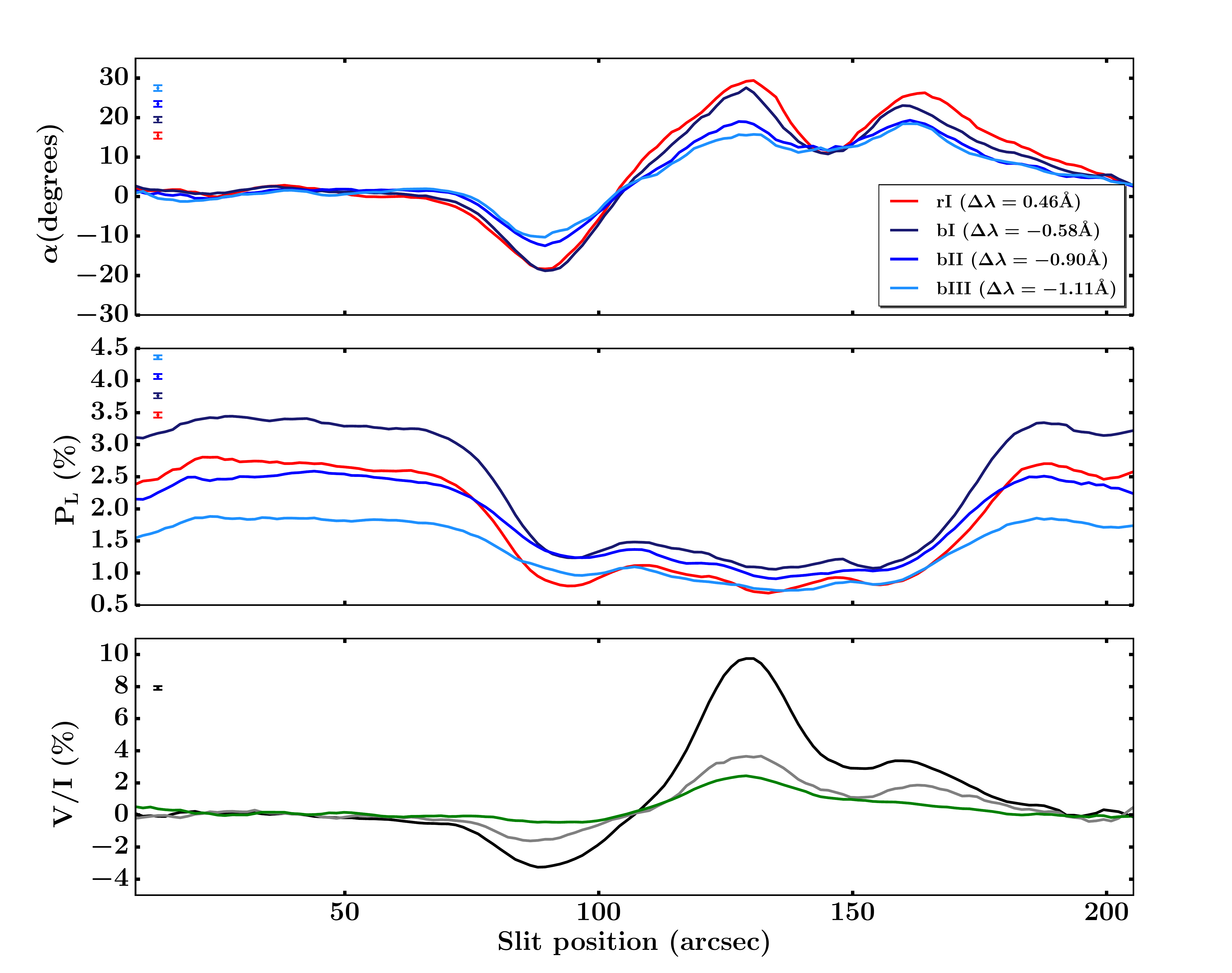}
        \caption{Linear polarization angle (top panel) and total linear polarization fraction (middle panel) as a function of the 
        position along the spectrograph's slit at four different wavelengths in the wings of the Ca~{\sc i} 
        line at 4227\,{\AA}. The various quantities are obtained by averaging the observed Stokes parameters over an 
        interval of 50\,m{\AA}, centered at the considered wavelength.  
        These wavelengths are referred to as rI (red), bI (dark blue), bII (blue), and bIII (cyan), as discussed in the main text. 
        The distance between the center of each interval and the center of the Ca~{\sc{i}} line ($\Delta \lambda = \lambda - \lambda_0$) is indicated in the legend. 
        For all figures the angle $\alpha$ is defined relative to a reference direction parallel to the nearest limb.
        In the bottom panel the amplitude of the blue peak of the $V/I$ signal of Fe~{\sc i} 4224.2\,{\AA} (black line), Fe~{\sc i} 
        4228.7\,{\AA} (green line), and Ca~{\sc i} 4227\,{\AA} (gray line) are shown as a function of the slit position.
        The error bars are reported in the upper left of each panel.
        These data refer to a spectropolarimetric observation of a moderately active region close to the west limb 
        performed on April 19, 2019. Figure adapted from \citet{Capozzi}.}
\label{fig_irsol}
\end{figure}
More recently, \cite{Alsina18} showed that, in addition, 
the scattering polarization wing lobes are sensitive to the magnetic fields in the photosphere. 
This sensitivity arises from the magneto-optical (MO) effects quantified by the 
anomalous dispersion coefficient $\rho_V$  entering the radiative transfer (RT) equation, which describes 
the coupling between the Stokes parameters $Q$ and $U$. 
Although the influence of MO effects in RT has been known for a long time, this had typically been observed and studied in the core 
of spectral lines, where their impact can be regarded as second-order effects, which are generally negligible except in the presence of strong 
magnetic fields. 
However, the theoretical works of \cite{Alsina16} and \cite{DelPinoAleman2016} showed that, in the line wings, $\rho_V$ becomes comparable 
to the absorption coefficient $\eta_I$ in the presence of magnetic fields with a relatively small longitudinal component (i.e., the projection of the magnetic field along the line of sight, LOS), with an intensity similar to those that 
characterize the onset of the Hanle effect \citep[see][]{Alsina19}.
If a sizable linear polarization signal is produced in the wings of the spectral line by another physical mechanism, most notably coherent scattering processes, 
it will then be strongly modified by such MO effects. This mechanism produces a rotation of the plane of linear polarization (also known  
as Faraday rotation), as well as a decrease in the degree of total linear polarization \citep{Alsina18}.\\
Throughout this work we do not consider magnetic fields of strengths typically found 
in active regions. Instead, we consider scenarios in which the splitting induced by the magnetic field in the atomic levels of the Ca~{\sc{i}} $4227$~{\AA} line is significantly smaller than the width of the line profile. In such cases the linear polarization signals produced by the Zeeman effect are relatively weak \citep[e.g., ][]{LandiE}. In the wings of the Ca~{\sc{i}} line their amplitudes 
are expected to be more than an order of magnitude smaller than that of the scattering polarization signals found close to the limb. For such field strengths, the $\rho_{Q}$ and $\rho_{U}$ anomalous dispersion coefficients, characterizing the MO effects that induce a coupling between linear and circular polarization states, are much smaller than the absorption coefficient $\eta_I$. Moreover, their ratios to $\eta_I$ sharply decrease when going from the line core into the line wings \citep[see][]{Alsina16,Alsina18}. Thus, in the following sections, the impact of these physical processes will be disregarded.

\section{\textbf{Observational finding}}
Figure~\ref{fig_irsol} shows the results of a spectropolarimetric observation of a region with a moderate level of magnetic activity located near the west limb of the solar disk, 
taken on April 19, 2019 \citep[see][]{Capozzi}.
The observation was performed with the Zurich Imaging Polarimeter \citep[ZIMPOL;][]{Ramelli10} 
at the Istituto Ricerche Solari Locarno (IRSOL).
The top and middle panels show the variation along the spectrograph’s slit of the linear polarization angle $\alpha$ and total linear polarization fraction $P_{L}$, respectively, at different wing wavelengths, corresponding to the local maxima in the wings of the observed $Q/I$ profiles
\citep[for details on the choice of the wavelengths, see][]{Capozzi}. The above-mentioned quantities are obtained by averaging the measured Stokes parameters over an interval 
of $50$ m\AA, 
centered around these wavelengths \citep[see][for more details]{Capozzi}. 
We refer to the considered wavelengths as rI (in the red wing of the line) and bI, bII, and bIII (in the blue wing, at increasing distance from line center). 
The distances from line center are reported in the legend of Fig.~\ref{fig_irsol}. 
In contrast, the theoretical values of the same quantities, presented in the following sections, refer to the same wavelengths, but 
without performing any average. 
We find this approach to be suitable, having verified numerically that the differences between the quantities obtained when selecting a specific wavelength and when performing the average over the above-mentioned  bandwidth are barely perceptible.

We recall that the linear polarization angle is the angle between the direction of linear polarization and a reference direction, which in this work is taken parallel to the nearest limb. 
The linear polarization angle is defined within the interval 
between $-90^\circ$ 
and $90^\circ$ 
\citep[e.g.,][]{LandiE}, and  is given by 
\begin{equation} 
\label{eqalpha} 
\alpha = \frac{1}{2} \tan^{-1} \left(\frac{U}{Q}\right) + \alpha_{0} \, , 
\end{equation}
with
\begin{equation}
\alpha_{0} = 
\begin{cases}
0 ^\circ,  \ \ \ \ \ & \text{if} \ Q > 0 \\
90 ^\circ,\ & \text{if} \ Q < 0 \ \text{and} \ U \geq 0 \quad . \\
-90 ^\circ,\ \ \ & \text{if} \ Q < 0 \ \text{and} \ U < 0  
\end{cases}
\end{equation}
This angle is positive when $U/I > 0$, meaning that the plane of linear polarization is rotated counter-clockwise 
relative to the above-mentioned reference direction. 
The total linear polarization fraction is given by
\begin{equation}\label{eqpl}
P_{L}= \sqrt{(Q/I)^2+(U/I)^2}.
\end{equation}

The bottom panel of Fig.~\ref{fig_irsol} shows the variation with the slit position of the amplitude of the blue peak of the $V/I$ signal 
of three different lines falling in the observed spectral 
interval, and forming in different atmospheric regions: Fe~{\sc i} 4228.7\,{\AA} (approximately mapping the middle photosphere, green curve),
Fe~{\sc i} 4224.2\,{\AA} (approximately mapping the upper photosphere, black curve), and Ca~{\sc i} 4227\,{\AA} (region between upper photosphere 
and lower chromosphere, gray curve).
Such spectral lines have effective Land\'e factors of $\bar{g} = 1.25$, $\bar{g} = 2.5$, and $\bar{g} = 1$, respectively. 
Information on the integration times for the above-mentioned observation, as well the uncertainties in the four Stokes 
parameters and in $\alpha$ and $P_L$ for each of the considered wavelengths,   
can be found in \cite{Capozzi}. 
The corresponding error bars are reported in  
the top left corner of each panel in Fig.~\ref{fig_irsol}. \\
In Fig.~\ref{fig_irsol} we highlight the appearance of three peaks, one negative and two positive, both in $\alpha$ (at the considered wing wavelengths) and in $V/I$. 
We recall that, for the relatively weak magnetic fields considered in this work, both these observables are
sensitive to the longitudinal component of the magnetic field via the MO (see above) and Zeeman \citep{LandiE} effects, respectively.   
The two positive peaks, 
found at the slit positions between $120\arcsec$ and $140\arcsec$ and 
between $155\arcsec$ and $175\arcsec$, show an interesting behavior.
For all three considered lines, the $V/I$ amplitude in the latter peak is substantially smaller than in the former, 
which 
is much more apparent for the lines forming in deeper layers (i.e., those corresponding to the black and green curves in the bottom 
panel).
By contrast, the two peaks in $\alpha$ 
reach similar values, despite also being mainly sensitive to photospheric magnetic fields.
The different behavior displayed by 
$\alpha$ and $V/I$ indicates that these two observables are sensitive to different properties of the magnetic field.
In $P_{L}$, we observe that its sharpest variations occur when going from the quieter regions (characterized by small $V/I$ and little variation in $\alpha$) to those with a higher level of activity. 
This suggests that, compared to $\alpha$, it is less sensitive to the magnetic field and more sensitive to other thermodynamic properties 
of the atmosphere, such as density and temperature.

\section{\textbf{Formulation and scope of the work}} \label{formulationscope}
In the present article  
we investigate the diagnostic potential of $\alpha$ in the wings of the Ca~{\sc i} 4227\,{\AA} line to infer information on the longitudinal component of photospheric magnetic fields. 

To this end, we numerically modeled the intensity and polarization profiles 
of this line by solving the 
RT problem for polarized radiation 
under non-local thermodynamic equilibrium (NLTE) conditions in a one-dimensional semi-empirical model of the solar atmosphere \citep[see][]{Alsina17}. 
Suitably modeling the broad wing linear polarization signals of this strong resonance line 
requires accounting for the frequency correlation between incoming and scattered radiation 
\citep[or partial frequency redistribution, PRD; e.g.,][]{Faurobert92}. Its influence must be taken into account together with that of the 
magnetic field, through 
the Hanle, Zeeman, and MO effects. The numerical approach considered in this work is based on a rigorous quantum theory of 
polarization 
that can jointly account for all such effects \citep{Bommier1997}.

Due to their strong influence on the broad wings of this line, our calculations also include the impact 
of a number of blended Fe~{\sc{i}} lines, which we modeled under the assumption of LTE, making use of the data presented in \cite{Kurucz}. 
We   verified numerically that the influence of such blends is small or marginal at wavelengths close to bI, bII, bIII, or rI, 
which are the focus of the present work. 
Thus, the accuracy of the RT modeling of such blended lines is not critical for the reliability of our results (see Appendix \ref{appendixA} for more details).

\section{Results} \label{results}
The synthetic profiles considered in this work are obtained from RT calculations, as described above, considering an LOS
with $\mu = \cos\theta = 0.1$, with $\theta$ the heliocentric angle, thus ensuring a large scattering polarization amplitude. For simplicity, we consider a magnetic field whose strength and orientation is constant at all heights in the atmospheric model. Recalling that the magnetic sensitivity in the wings is primarily due to the MO effects that are proportional to the longitudinal component of the magnetic field \citep{Alsina18}, the calculations presented in this work consider horizontal magnetic fields contained in the plane defined by the vertical and the LOS. 

\subsection{Volume-filling magnetic field}
\label{sec:volumefilling}
\begin{figure*}[h!]
\centering
\includegraphics[width=.7\textwidth]{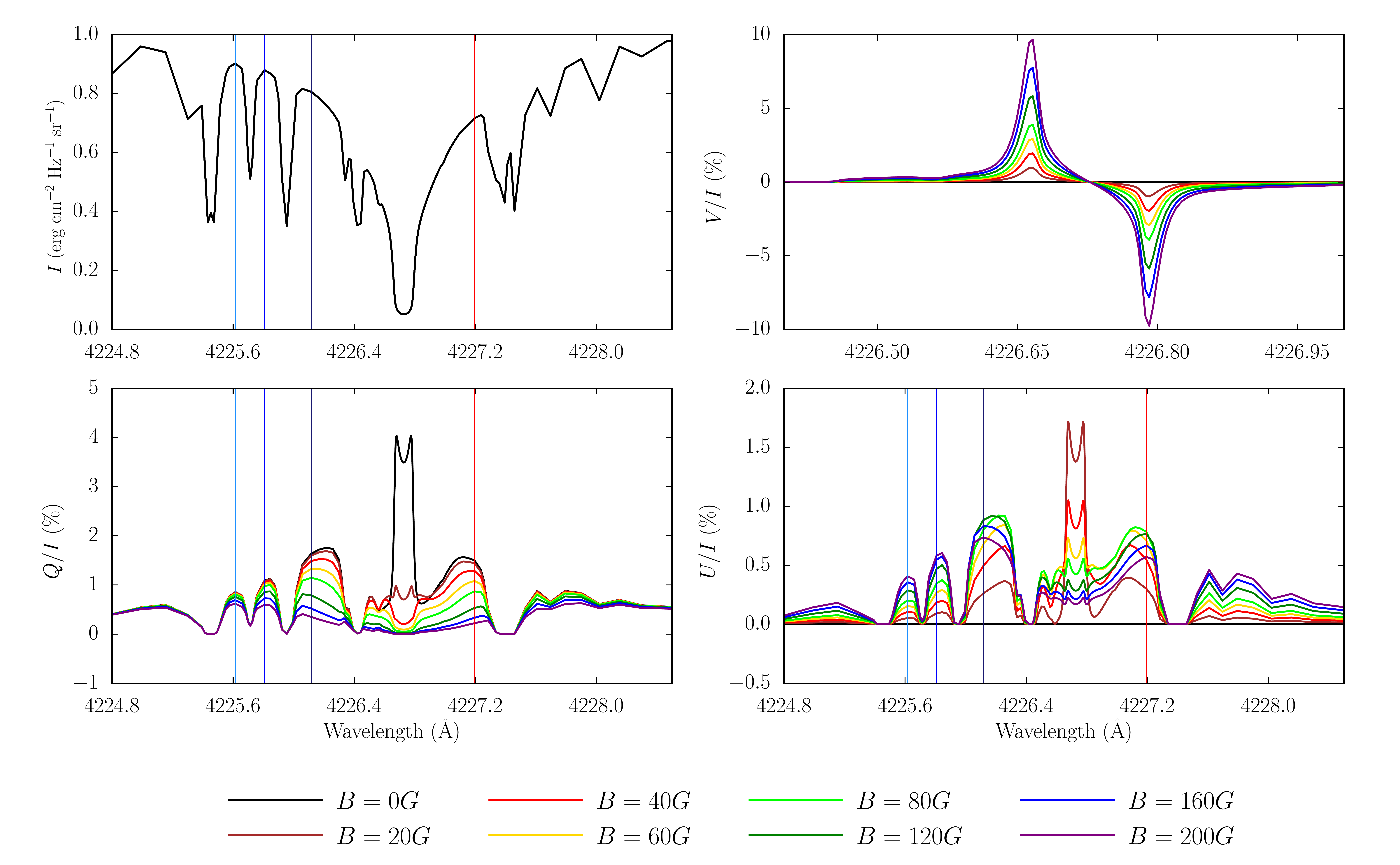}
        \caption{Stokes $I$, $V/I$, $Q/I$, and $U/I$ profiles of 
        the Ca~{\sc i} 4227\,{\AA} obtained 
        through the RT calculations described in the text. In this figure and those shown below, a LOS with $\mu = 0.1$ is considered, and the magnetic 
        fields are taken to be horizontal and contained in the plane defined by the vertical and the LOS. 
        The FAL-C atmospheric model has been used for the calculations and field strengths between $0$\,G and $200$\,G are considered (see legend). 
    The vertical lines indicate the wavelengths 
    introduced in \cite{Capozzi},
    corresponding to rI (red), 
    bI (dark blue), bII (blue), and bIII (cyan). 
    A narrower spectral range is considered for the panel showing the Stokes $V/I$ profile. The positive reference direction for Stokes $Q$ is taken parallel to the nearest limb.}
\label{plot_stok_B_barklem}
\end{figure*}

In this section we consider the case of a magnetic field that occupies the entirety of the  resolution element, which we hereafter refer to as volume-filling.
The sensitivity of the scattering polarization wings of Ca~{\sc i} $4227$ 
to such magnetic fields, calculated using the the semi-empirical model C of \citet{Fontenla93}, hereafter FAL-C,
is illustrated in 
Fig.~\ref{plot_stok_B_barklem}. 
This figure shows that even relatively weak magnetic fields give rise to signals of considerable amplitude in the wings of the $U/I$ profile and a conspicuous magnetic sensitivity in both the $Q/I$ and $U/I$ wings.
Furthermore, as the magnetic field increases an overall decrease in the linear polarization 
fraction can also be clearly observed \citep[for a more in-depth discussion, see][]{Alsina18}.
The magnetic sensitivity found in the line-core scattering polarization signal at similar field strengths is due to the Hanle effect. 
On the other hand, the circular polarization $V/I$ signal found around the core region is produced by the Zeeman effect.

The fact that the magnetic sensitivity of the various Stokes parameters at different wavelengths is driven mainly by different physical mechanisms is supported by compelling theoretical arguments \citep{LandiE,Sampoorna09,Alsina16}.
For the Ca~{\sc{i}} line considered here, we  verified this through a numerical investigation based on response functions.
The same investigation confirms that
the magnetic sensitivity in the wing scattering polarization arises from MO effects, primarily due to the fields present in 
a narrow range of photospheric depths, extending roughly $300$~km (see Appendix \ref{appendixB} for more details). 

\subsection{Magnetic fields occupying a fraction of the resolution element} \label{sec::ResultsTwoComp}
In general, the magnetic field may permeate only a given fraction of the resolution element. 
The observed radiation results from the contribution from all spatial points within this area, each with its own magnetic field strength and orientation. 
It is well known that the circular polarization signals produced by the Zeeman effect are sensitive to the magnetic flux, but not to the fraction of the area that the magnetic field occupies. 
On the other hand, previous theoretical investigations \citep{Alsina19}
hinted that the $\alpha$ value in the line wings may depend on this fraction and  on the flux. 
In order to quantify this sensitivity, we assume that a
uniform magnetic field is present in a fraction of the resolution element, 
whereas the rest is devoid of any magnetic field.
We compare several 
realizations in which magnetic fields with different strength $B$ are present within a different fraction $f$ of the total area (or filling factor). 
These values were selected so that the average magnetic field over the resolution element, given by
\begin{equation}
        B_{\mbox{\scriptsize av}} = f \, B \, ,
\end{equation}
which is proportional to the flux,\footnote{Observing that the longitudinal component of the considered magnetic field is directed towards the observer, the average field and the flux are defined as positive.}
is the same for all cases.
Specifically, we assumed $f=1$ and $B=40$\,G (representative of a volume-filling magnetic field), $f = 0.5$ and $B = 80$\,G, and
$f = 0.2$ and $B = 200$\,G, all corresponding to $B_{\mbox{\scriptsize av}}=40$\,G. 

\begin{figure*}
\centering
\includegraphics[width=0.8\textwidth]{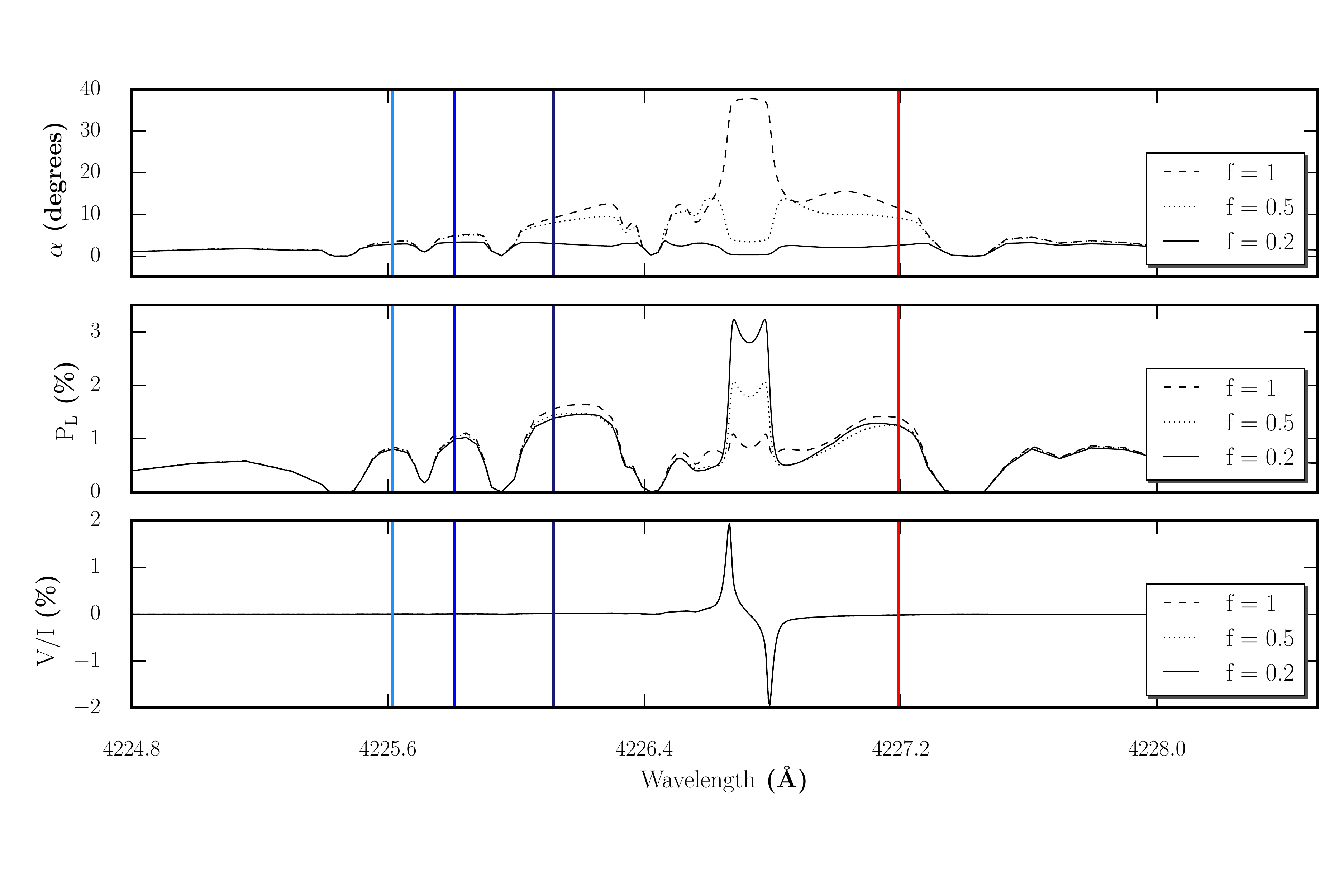}
        \caption{
        Linear polarization angle $\alpha$ 
        (top panel), linear polarization fraction $P_L$ (middle 
        panel), and Stokes $V/I$ (bottom panel) across the Ca~{\sc i}
        4227\,\AA\ line, calculated in the FAL-C 
        atmospheric model for an LOS with $\mu=0.1$. 
        The various curves correspond to calculations 
        considering horizontal fields with 
        $B_{\mathrm{av}} = 40$\,G (see text), 
        but different filling factors and field strengths: $f=1$ and $B=40$\,G (dashed line), $f=0.5$ and $B=80$\,G (dotted line), and $f = 0.2$ and $B=200$\,G (solid line). The vertical lines indicate the same wavelengths as in the previous figure.} 
\label{promedioValpha}
\end{figure*}

The observables $\alpha$, $P_{L}$, and $V/I$ resulting from these realizations are shown in the top, middle, and bottom panel, respectively, of Fig.~\ref{promedioValpha}. 
The Stokes $V/I$ signals found in the core of the Ca~{\sc{i}} line coincide for the three considered cases. 
We note that the RT code used in this investigation does not allow modeling the $V/I$ signals in the nearby 
blended lines, and that the purpose of presenting such a comparison is only to emphasize that
all the circular polarization signals produced by the Zeeman effect  
depend primarily on
the magnetic flux 
when considering relatively weak magnetic fields as in the present work
\citep[e.g.,][]{LandiE}.
By contrast, $\alpha$ and $P_{L}$ present clear differences between the three realizations, which can be found both in the line core, where they can be attributed to the impact of the Hanle effect, and in the line wings, where they are due to MO effects. 
\begin{figure*}[t!]
\centering
\includegraphics[width=0.43\textwidth]{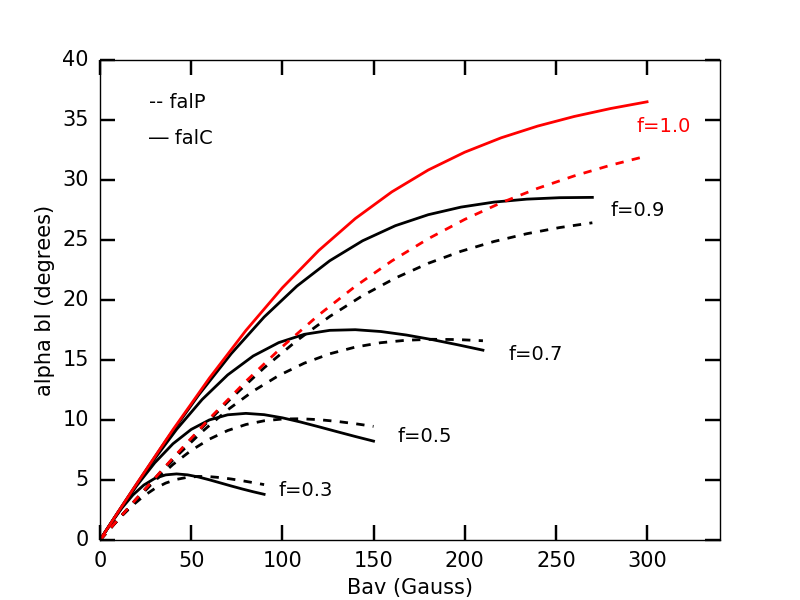}
\includegraphics[width=0.43\textwidth]{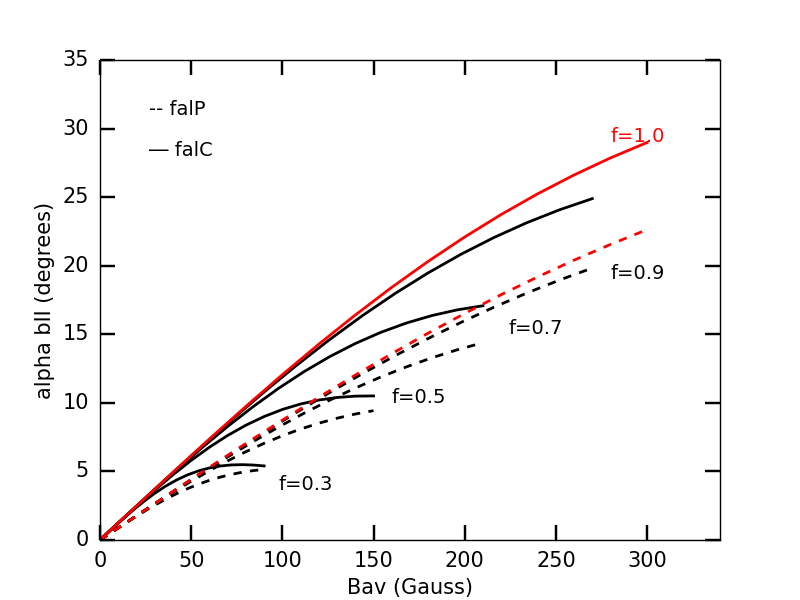}
\caption{Linear polarization angle $\alpha$ at the bI (left panel) and bII (right panel) 
    wavelengths (see text) for an LOS with $\mu=0.1$
    as a function of 
        $B_{\mbox{\scriptsize av}}$ considering horizontal magnetic fields with 
        different filling factors $f$, for the atmospheric models FAL-C (solid line) and  
        FAL-P (dashed line). For the sake of clarity, the curves corresponding to 
        $f = 1$ are shown in red.} 
\label{fig:angle_vs_B_noffmeno}
\end{figure*}

The magnetic sensitivity of the line wings is further highlighted by
the results shown in Fig.~\ref{fig:angle_vs_B_noffmeno}, where the polarization angles at the 
bI (left panel) and bII (right panel) wing wavelengths are plotted as a function of $B_{\mathrm{av}}$, considering horizontal magnetic fields with different filling factors. 
For each considered factor $f$ (see legend) we took a series of realizations in which the field strength within the magnetized region $B$ increases from $0$ to $300$~G in steps of $20$~G, considering also the $10$~G case. 
All such calculations were carried out considering both models C and P of \citet{Fontenla93}, representative of typical conditions of an average 
quiet region of the solar atmosphere and of a solar plage, respectively.
For a given value of $B_{\mbox{\scriptsize av}}$, the polarization angle strongly depends on the value of $f$, sharply increasing with it.
This is a finding of great
diagnostic interest as it implies 
that the values of $\alpha$ at such wavelengths encode information on the fraction of resolution element occupied by the magnetic field, 
which can be directly obtained  
if an independent estimate of the magnetic flux is available.
Such information may be accessible, for instance, from the circular polarization signals of photospheric spectral lines,  
and could be determined through the application of a number of widely known techniques, including 
for example the magnetograph formula or inversion codes that are freely available to the solar physics 
community \citep[e.g.,][]{RuizCoboDelToroIniesta92,SocasNavarro+15}. 

We observe that this proposed diagnostic approach 
relies on the assumption that the strength and orientation of the magnetic field is constant with height in the solar atmosphere, 
similar to other commonly used diagnostic methods such as the magnetograph formula. In this case this assumption is justified on 
the grounds that 
the scattering polarization wings of the Ca~{\sc{i}} are sensitive, via MO effects, to the magnetic fields in a relatively narrow 
range of photospheric heights.

From Fig.~\ref{fig:angle_vs_B_noffmeno} it can be seen that the sensitivity of $\alpha$ with respect to $f$ is higher at the bI wavelength  
than at  bII. This is a consequence of the larger impact of MO effects at wing wavelengths closer to the line center. 
We also verified that this sensitivity is even lower at the bIII wavelength, while those at rI and bI are very similar to each other. 
The results of Fig.~\ref{fig:angle_vs_B_noffmeno} also show that the magnetic sensitivity of $\alpha$ is appreciably model-dependent.  
Thus, in order to reliably use this observable to infer the magnetic filling factor, one must 
also obtain information about thermodynamic properties of the atmosphere, for instance, by making use of the above-mentioned inversion codes.  

Furthermore, we find the polarization fraction $P_{L}$ (not shown in the previous figure) to be far less suitable as an observable 
for extracting information on the magnetic field. 
By contrast to $\alpha$, which is zero in the absence of a magnetic field, its zero-field reference value is not known 
because it depends primarily on the radiation anisotropy in the solar atmosphere.
For this reason, it is far more model-dependent than $\alpha$, and therefore very precise information on the temperature and density
stratification of the solar atmosphere would be necessary in order to reliably use it for magnetic diagnostics.

\section{Conclusions}
The present work builds upon the theoretical finding that the scattering polarization wings of strong resonance lines are sensitive to the longitudinal component of relatively weak magnetic fields through MO effects.
For the Ca~{\sc i} 4227\,\AA\ line, this mechanism allows the investigation of photospheric magnetic fields 
in the gauss range, in which the Hanle effect also operates.

Further motivated by recent spectropolarimetric observations \citep{Capozzi}, 
the present investigation shows that the linear polarization angle $\alpha$ measured in the wings of the Ca~{\sc i} 4227\,{\AA} line encodes information not only on the magnetic flux, but also on the filling factor of the magnetic field, which remains hidden to diagnostic techniques based on the circular polarization signals produced by the Zeeman effect. 
This information can be directly obtained from $\alpha$ if the magnetic flux is known (e.g., by applying well-established Zeeman-based methods
which rely on the weak-field assumption). 
Observations of this line with reasonably short integration times \citep[see][]{Capozzi} yield relatively low uncertainties for 
$\alpha$ in the wings, suggesting that such diagnostic methods should be suitable for applying valuable constraints on the filling factor of photospheric magnetic fields. 
Furthermore, observing that the core of the Ca~{\sc i} 4227\,{\AA} line forms in the lower chromosphere, 
the magnetic sensitivity of the scattering polarization in the wings (due to MO effects) can be exploited in combination with that in the core (due to the Hanle and Zeeman effects) 
to simultaneously obtain
information on the magnetic fields present at different atmospheric heights, from the photosphere to the chromosphere. 

The results of this work can also be applied to other strong resonance lines with extended wing scattering polarization signals. 
Remarkable examples are the H~{\sc i} Ly-$\alpha$ line at 1215\,{\AA} \citep{Alsina19}, and the Mg~{\sc ii} k line at 2795\,{\AA} 
\citep{Alsina16,DelPinoAleman2016,delPino+20}, whose near wings originate at chromospheric heights. 
The magnetic sensitivity of $\alpha$ is necessarily model-dependent. The required information on the thermodynamic properties of 
the observed atmospheric region may be accessible through the application of state-of-the-art inversion codes. 
The findings shown in this article
illustrate that new inversion codes fully accounting for the impact of magnetic fields of arbitrary strength and orientation in strong resonance lines, combining the Hanle, Zeeman, and MO effects, would be very valuable tools to access unexplored aspects of solar magnetism.  

\begin{acknowledgements}
        This research work was financed by the Swiss National Science Foundation (SNSF) through grants 200020\_169418
        and 200020\_184952.
        E.A.B. and L.B. gratefully acknowledge financial support 
        by SNSF through grant 200021\_175997. 
        L.B. and J.T.B. gratefully acknowledge financial support by SNSF through grant CRSII5\_180238. 
        J.T.B. acknowledges the funding received from the European Research Council through 
        Advanced Grant Agreement no. 742265. 
        IRSOL is supported by the Swiss Confederation (SEFRI), Canton Ticino, 
        the city of Locarno and the local municipalities. 
\end{acknowledgements}

\bibliographystyle{aa}
\bibliography{capozzi2021}
\onecolumn
\appendix 


\section{\textbf{Impact of the blends on the polarization angle}}\label{appendixA} 

The calculations presented in this article include the impact of a number of blended Fe~{\sc{i}} lines
(see Sect.~\ref{formulationscope}).
Given that these Fe~{\sc i} lines are relatively weak, and that their NLTE modeling would represent a 
major task, we modeled them under the assumption of LTE. 
Their impact was taken into account by adding at each point in the spectral grid of the RT calculations 
their opacity and emissivity contributions to those of the continuum. 
Here, we analyze the influence of these blends on the scattering polarization profiles through RT calculations like those presented in the main text of the article, considering the FAL-C atmospheric model, for an LOS with $\mu=0.1$. 
In this analysis we focus particularly on the reference wavelength positions introduced in the main text (bI, bII, bIII, and rI).
\begin{figure}[h!]
\centering
\includegraphics[width=0.7\textwidth,clip=]{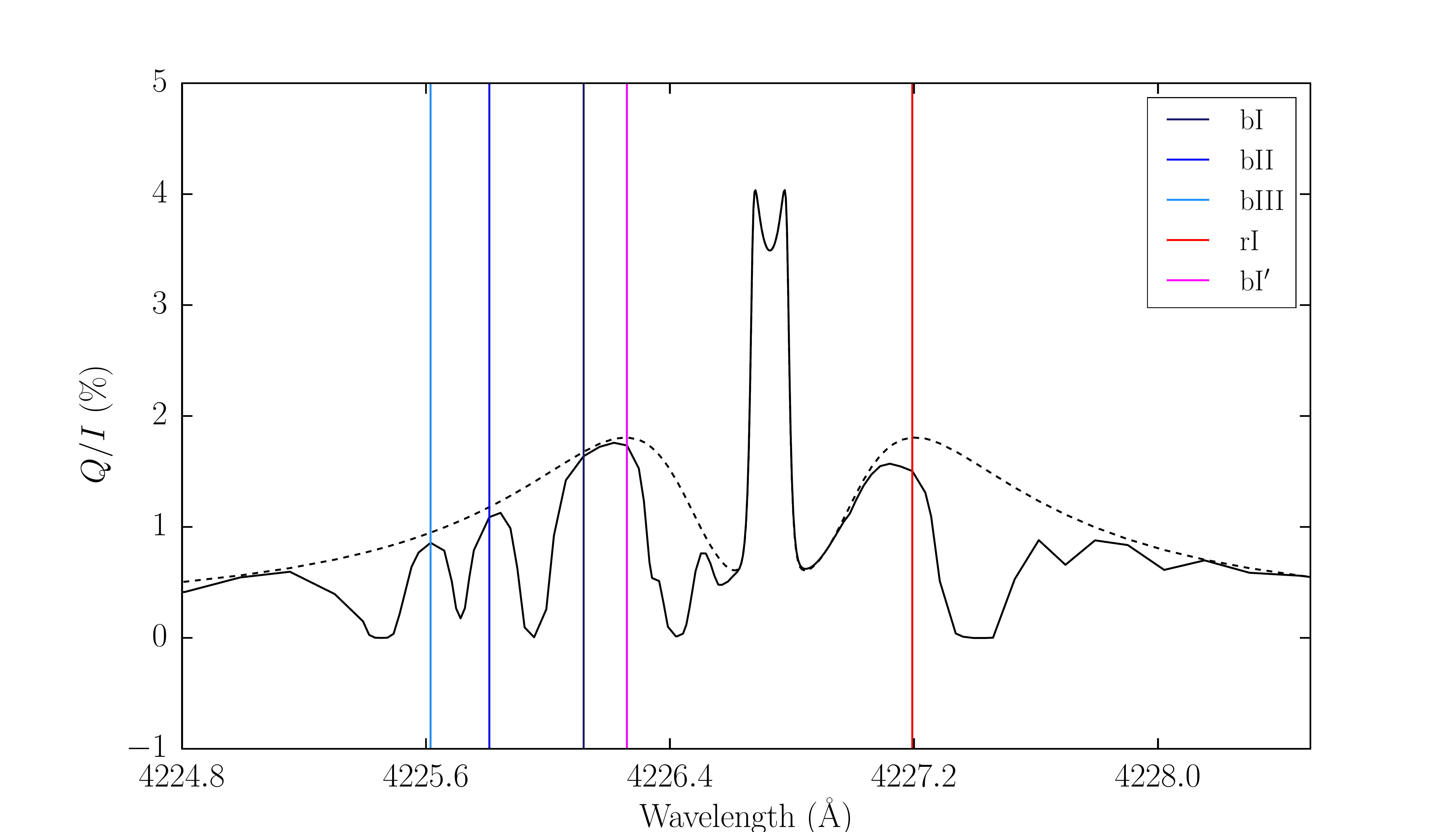}
 \hspace*{-0.03\textwidth}
\includegraphics[width=0.9\textwidth]{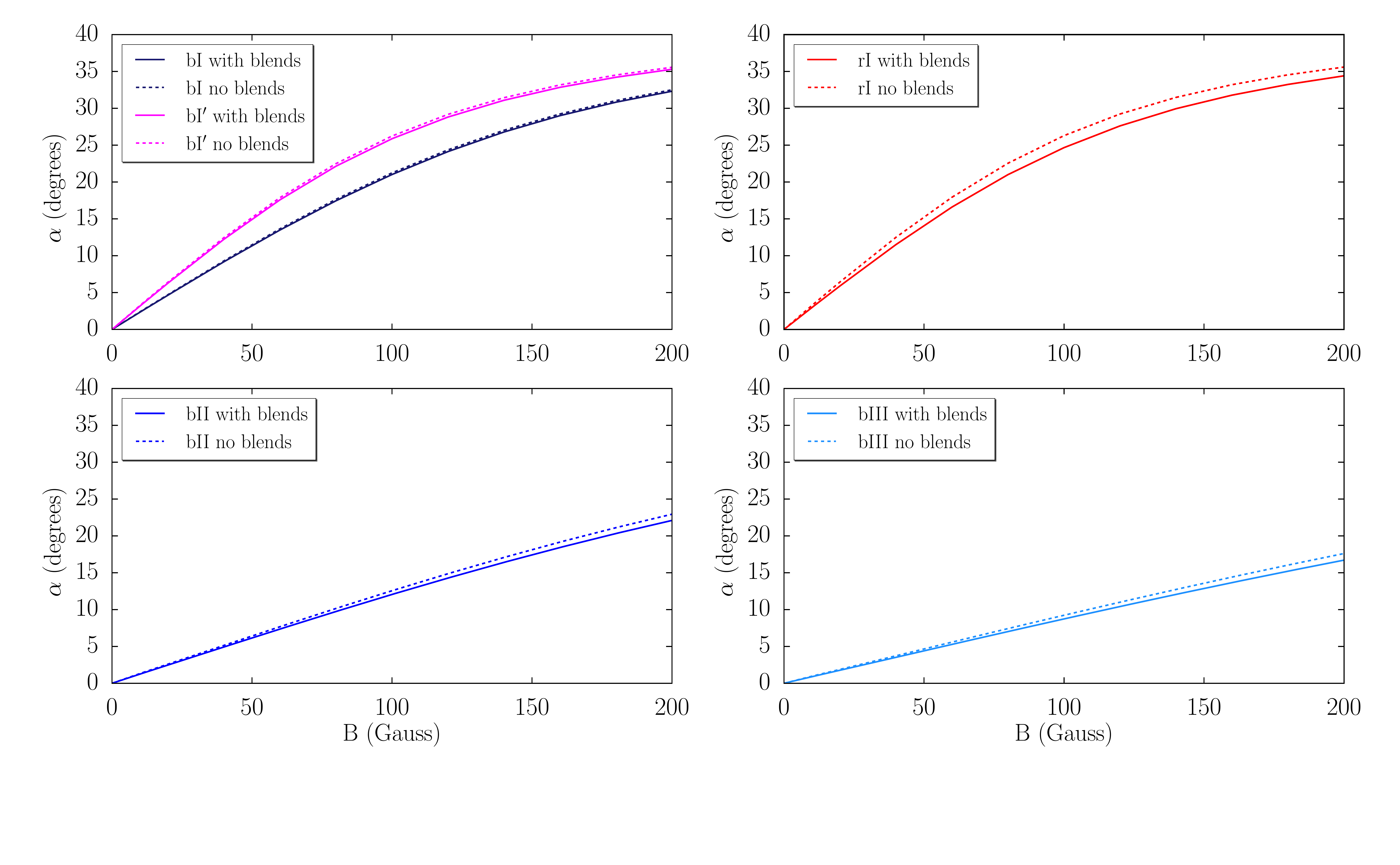}
 \vspace{0.005\textwidth}
        \caption{Results of RT calculations considering the FAL-C atmospheric model, for the radiation emerging at a LOS with $\mu=0.1$.
        The top panel shows the $Q/I$ profile of the Ca~{\sc i} 4227\,\AA\ line 
        as a function of wavelength, obtained in the absence of magnetic fields.
        The solid curve shows the profile obtained when including the effects of blends, and the dashed curve shows the profile obtained when neglecting them. 
        The colored vertical lines give the wavelength positions of interest introduced in the article (see legend), with bI$^{\prime}$ defined in the text. 
        The bottom panels show the linear polarization angle $\alpha$ as a function of field strength, at the various wavelength positions of interest (see legend). 
        These calculations were performed in the presence of a horizontal magnetic field contained in the plane defined by the local vertical and the LOS (see text). 
        The solid and dashed curves represent the results obtained when accounting for the blends and neglecting them, respectively.}    
\label{angle_bnob}
\end{figure}
The upper panel of Figure~\ref{angle_bnob} shows the $Q/I$ 
profiles calculated in the absence of magnetic fields, comparing the results obtained when taking into account (solid curve) and neglecting the blends (dashed curve). For all the wavelengths of interest, highlighted with colored vertical lines (see legend), the blends appear to have a marginal impact, with the exception of rI. 
In the same panel, 
the magenta line indicates the wavelength position in the blue wing where the linear polarization amplitude is largest when all blends are neglected (hereafter bI$^{\prime}$).

The four lower panels show the variation of the linear polarization angle $\alpha$ with 
the magnetic field strength for the above-mentioned wavelengths, 
accounting for (solid curves) and neglecting (dashed curves) 
the impact of blends.
As in the main text, the orientation of the magnetic fields is taken to be horizontal and contained in the plane defined by the LOS and the local 
vertical. The RT calculations were carried out for magnetic field strengths between $0$ and $200$~G in increments of $1$~G. 
This figure clearly shows that the inclusion of blends has comparatively little impact on both the value of 
$\alpha$ and its magnetic dependence (compare dashed and solid line in the figure), even for rI. 
This supports the claim made in the main text that the accuracy in the modeling of the blends is not critical for the reliability of our RT calculations. 
In agreement with the results shown in  Fig.~\ref{fig:angle_vs_B_noffmeno}, for all considered wavelengths, we find a monotonic increase of $\alpha$ with field 
strength, eventually reaching a plateau.  
We observe that such magnetic sensitivity depends on the considered wavelength, being stronger for those closer to the line core. 


\section{\textbf{Response function}}\label{appendixB}
As pointed out in Sect.~\ref{results}, depending on the considered spectral region, 
the magnetic sensitivity of the various Stokes parameters of the Ca~{\sc{i}} line is driven by different physical mechanisms.
Here, this is illustrated through a numerical investigation based on response functions. 
The response functions \citep[RFs, see][]{RFB,RFA,Uitenbroek2006} 
characterize the response of the intensity and polarization of the emergent radiation to the variation of a given physical quantity at any height and for a given wavelength.  
The RFs for the magnetic field for each of the Stokes parameters $S = \bigl\{I, Q, U, V \bigr\}$ can be 
determined numerically through
\begin{equation}\label{eqRF} 
 R_\lambda^{S,B}(z) = \frac{1}{\Delta B} \frac{\mathrm{d}}{\mathrm{d}z} \Delta S^{(z)}_\lambda \, ,
\end{equation}
where $\Delta S_\lambda^{(z)}=S_\lambda^{(z)}-S_\lambda^{(0)}$, with $S_\lambda^{(z)}$ and $S_\lambda^{(0)}$ the Stokes parameters at wavelength $\lambda$ obtained 
when applying a perturbation up to height point $z$ and without applying it, respectively. 
These Stokes parameters are obtained from RT calculations using atmospheric model FAL-C and for an LOS with $\mu = 0.1$.
We took the perturbation to be a $20$~G horizontal magnetic field contained in the plane defined by the LOS and the local vertical, 
thus having $\Delta B = 20$~G in Eq.~(\ref{eqRF}).
For $S_\lambda^{(z)}$ this perturbation is applied form the lower boundary of the considered atmospheric model 
up to the spatial grid point corresponding to height $z$. 
In such calculations the impact of the blended lines was taken into account (see Sect.~\ref{formulationscope} and Appendix~\ref{appendixA} for details).

\begin{figure}[h!]
\centering
\includegraphics[width=0.8\textwidth]{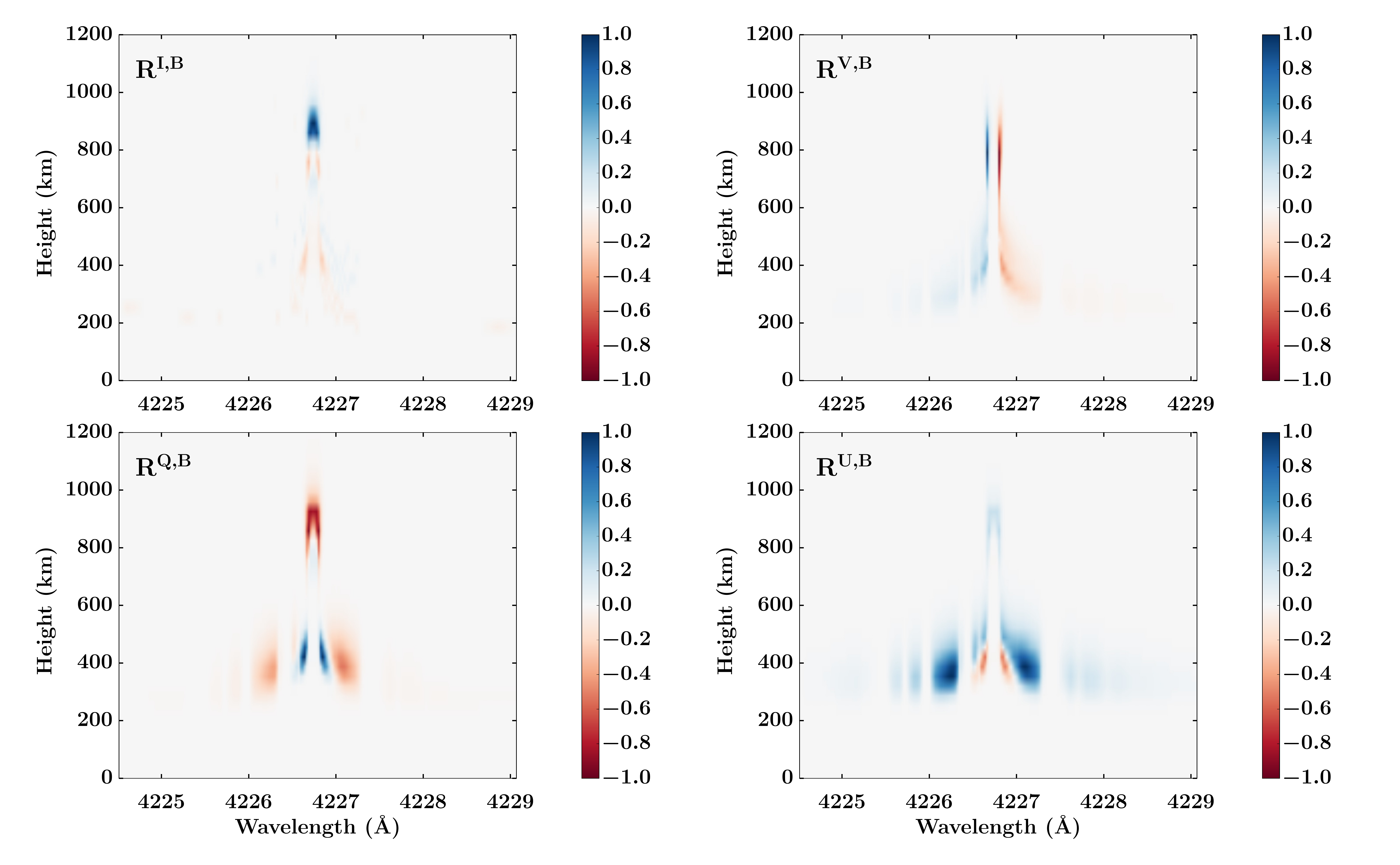}
\includegraphics[width=0.8\textwidth]{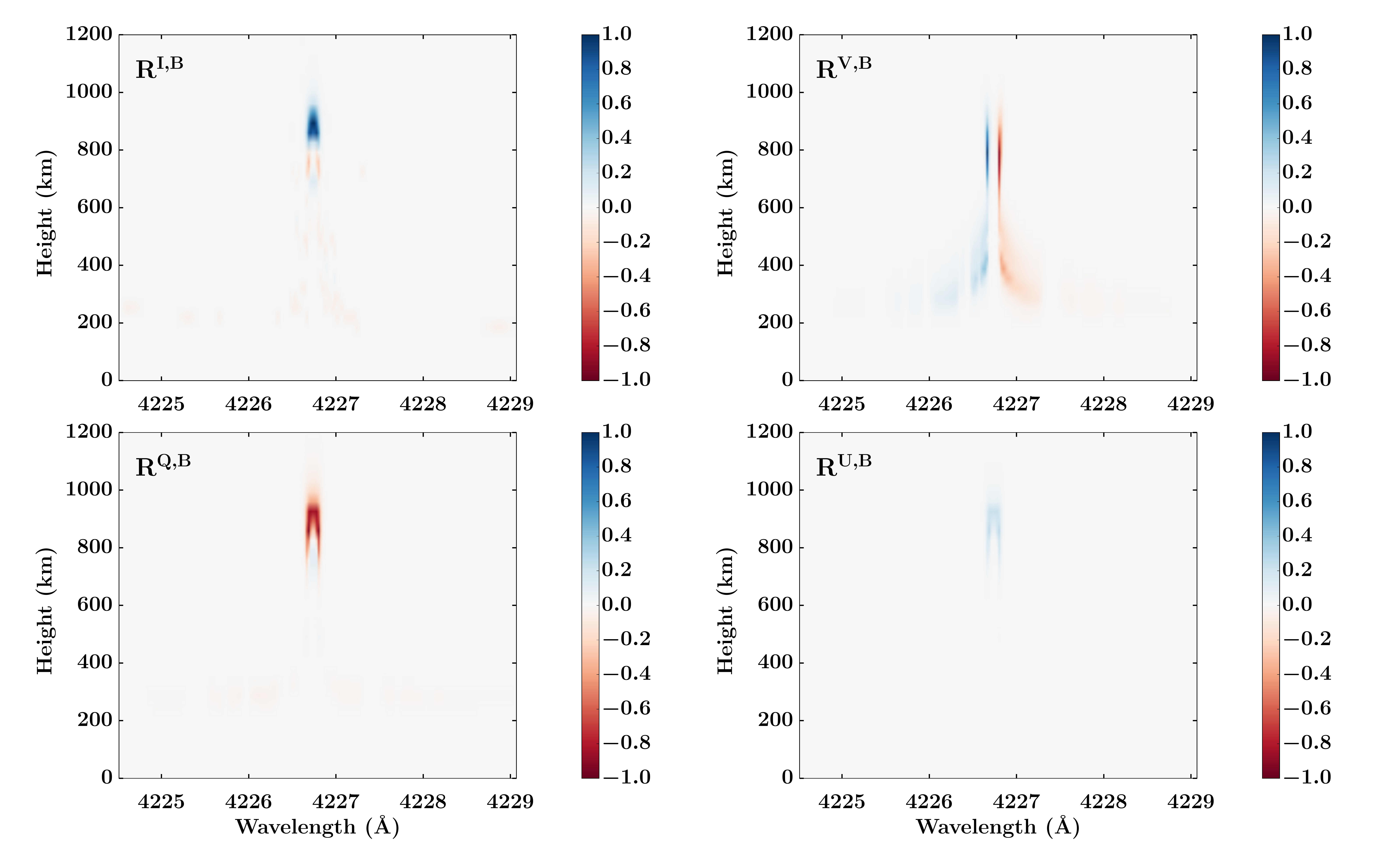}
  \caption{Response function of the Ca~{\sc i} 4227\,\AA\ line, calculated in the FAL-C atmospheric model 
  for an LOS with $\mu=0.1$. 
  The upper four   panels show,  clockwise   from the upper left, the RF for the Stokes parameters $I$, $V$, $U$, and $Q$, accounting 
  for all MO effects.
  The  bottom four panels show the RF for the same Stokes parameters, artificially setting the $\rho_{V}$ coefficients to zero in the RT 
  calculations from which the RF is computed.} 
\label{rfunctionMO_NOMO}
\end{figure}

To highlight the impact of Faraday rotation in the wing linear polarization signals, we computed the RFs both 
accounting for all MO effects in the spectral synthesis (see the  upper four panels of Fig.~\ref{rfunctionMO_NOMO}) and  
artificially setting to zero the anomalous dispersion coefficient $\rho_V$ in the transfer equation (see the lower four panels of 
Fig.~\ref{rfunctionMO_NOMO}). 
In order to ease this comparison, the RF for each Stokes parameter was normalized to its maximum absolute value for all heights and frequencies, 
considering both the calculations accounting for $\rho_V$ and neglecting it\footnote{Because of this normalization the 
RFs indicate the heights and frequencies where the impact of the magnetic field is strongest, 
but not the absolute value of the variation in the signal due to the considered perturbation.}. 
It can be observed that the intensity and the linear polarization signals in the line core are most sensitive to variations in the magnetic field at depths corresponding to the middle chromosphere (between $800$ and $1000$ km). 
The magnetic sensitivity of the circular polarization signals, at the wavelength positions corresponding to their maxima, is strongest at 
slightly greater depths, 
centered at roughly  $800$~km.
These results are in agreement with those of \citet{Supriya}. 

Farther into the line wings the magnetic sensitivity of the four Stokes parameters is strongest at considerably deeper layers of the solar atmosphere.
The comparison between the RFs for Stokes $Q$ and $U$ obtained accounting for the above-mentioned MO effects 
and neglecting them 
indicates that 
the strong sensitivity in the wing linear polarization to magnetic fields at photospheric heights of around $300$ km is almost entirely a consequence of such effects.
In the wings the strongest response of the circular polarization is found at heights that correspond 
to a range that goes from the photosphere to the lower chromosphere. 
As wing wavelengths farther from the line core are considered, this sensitivity both weakens and is centered on slightly deeper regions.

\end{document}